\documentclass{article}%
\usepackage{amsmath}
\usepackage{amsfonts}
\usepackage{amssymb}
\usepackage{graphicx}%
\providecommand{\U}[1]{\protect\rule{.1in}{.1in}}
\newtheorem{theorem}{Theorem}

\newtheorem{definition}[theorem]{Definition}

\begin{document}

\title{Wigner's infinite spin representations and inert matter\\{\small dedicated to the memory of Robert Schrader}}
\author{Bert Schroer\\present address: CBPF, Rua Dr. Xavier Sigaud 150, \\22290-180 Rio de Janeiro, Brazil\\permanent address: Institut f\"{u}r Theoretische Physik\\FU-Berlin, Arnimallee 14, 14195 Berlin, Germany}
\date{September 2015}
\maketitle

\begin{abstract}
Positive energy ray representations of the Poincar\'{e} group are naturally
subdivided into three classes according to their mass and spin content: m%
$>$%
0, m=0 finite helicity and m=0 infinite helicity. For a long time the
localization properties of the massless infinite spin class remained unknown
before it became clear that such matter does not permit compact spactime
localization and its generating covariant fields are localized on
semi-infinite spacelike strings.

Using a new perturbation theory for higher spin fields we show that infinite
spin matter cannot interact with normal matter and we formulate condition
under which this also could happen for finite spin $s>1$ fields. This raises
the question of a possible connection between inert matter and dark matter.

\end{abstract}

\section{Wigner's infinite spin representation and string-localization}

Wigner's famous 1939 theory of unitary representations of the Poincar\'{e}
group $\mathcal{P}$ was the first systematic and successful attempt to
classify relativistic particles according to the \textit{intrinsic} principles
of relativistic quantum theory \cite{Wig}. As we know nowadays, his massive
and massless spin/helicity class of positive energy ray representations of
$\mathcal{P}$ does not only cover all known particles, but their
"covariantization" \cite{Wein} leads also to a complete description of all
covariant point-local free fields. For each with physical spin or helicity
compatible covariant transformation property there exists a point-local (pl) field.

The only presently known way to describe interactions in four-dimensional
Minkowski space is to start from a scalar interaction density in terms of
Wick-products of free fields with the lowest short distance dimension and use
it as the starting point of the cutoff- and regularization-free causal
perturbation theory \cite{E-G}. These free fields do not have to be
Euler-Lagrange fields; perturbative QFT can be fully accounted for in terms of
interaction densities defined in terms of free fields obtained from Wigner's
representation theory without referring to any classical parallelism.

All positive energy representations are "induced" from irreducible
representations of the "little group". This subgroup of the Lorentz group is
the stability group of a conveniently chosen reference momentum on the forward
mass shell $H_{+}~$respectively the forward surface of the light cone$V_{+}$.
For $m>0~$this is a rotation subgroup of the Lorentz group and for $m=0~$the
noncompact Euclidean subgroup $E(2).$ Whereas the massive representation class
$(m>0,s=\frac{n}{2})$, of particles with mass $m$ and spin $s~$covers all
known massive particles (the first Wigner class), the massless representations
split into two quite different classes.

For the finite helicity representations the $E(2)~$subgroup of
Lorentz-"translations" are trivially represented ("degenerate"
representations), so that only the abelian rotation subgroup $U(1)\subset
E(2)$ remains; this accounts for the semi-integer helicity $\pm\left\vert
h\right\vert ,\left\vert h\right\vert =\frac{n}{2}$ representations (the
second Wigner class). The third Wigner class consists of \textit{faithful}
unitary representations of $E(2).$ Being a noncompact group, they are
necessarily \textit{infinite dimensional} and their irreducible components are
characterized in terms of a continuous Pauli-Lubanski invariant $\kappa.$

Since this invariant for massive representations is related to the spin as
$\kappa^{2}=m^{2}s(s+1)$ one may at first think that the properties of this
infinite spin matter can be studied by considering it as a limit
$m\rightarrow0,s\rightarrow\infty$ with $\kappa$ fixed. However it turns out
that the $(m,s)$ spinorial fields do not possess such an infinite spin limit.
Our main result concerning the \textit{impossibility of quantum field
theoretical interactions between WS with normal matter} depends among other
things on the absence of such an approximation; for this reason we will refer
to these representations briefly as the "Wigner stuff" (WS). This terminology
is also intended to highlight some of the mystery which surrounded this class
for the more than 6 decades after its discovery and which also the present
paper does not fully remove.

For a long time the WS representation class did not reveal its quantum field
theoretic localization properties. The standard group theoretical
covariantization method to construct intertwiners \cite{Wein} which convert
Wigner's unitary representations into their associated pl quantum fields does
not work for the WS representations.\ Hence it is not surprising that attempts
in \cite{A} (and more recently in \cite{Schu}) which aim at the construction
of covariant wave functions and associated Lagrangians fell short of solving
the issue of localization. In fact an important theorem \cite{Y} dating back
to the 70s proved that it is not possible to associate pl fields (Wightman
fields) with these representations.

Using the concept of modular localization, Brunetti, Guido and Longo showed
that WS representations permit to construct subspaces which are "modular
localized" in arbitrary narrow spacelike cones \cite{BGL} whose core is a
semi-infinite string. In subsequent work \cite{MSY} \cite{Rio} such generating
string-local covariant fields were constructed in terms of modular
localization concepts. In the same paper attempts were undertaken to show that
such string-local fields cannot have pl composites. These considerations were
strengthened in \cite{Koe}. A rigorous proof which excludes the possibility of
finding compact localized subalgebras (related to testfunction-smeared pl
fields) was finally presented in an seminal paper by Longo, Morinelli and
Rehren \cite{LMR}.

Being a positive energy representation, WS shares its stability property and
its ability to couple to gravity through its energy-momentum tensor with the
other two positive energy classes; hence it cannot be dismissed from the
outset as being unphysical. It will be shown that this form of matter is
\textit{inert} which means that it cannot interact with normal matter. There
are reason to believe that higher spin fields $s\geq2\ $may share this lack of
reactivity, but the additional calculations which would be necessary to
resolve this problem are outside the scope of this paper.

Absence or at least reduced reactivity is also a property of the ubiquitous
dark matter; only additional measurement in underground counters will be able
to resolve this problem.

The aim of the present paper is to convert the question of whether nature uses
WS matter into a problem of particle theory. Since the task of local quantum
physics (LQP) is to explain properties of matter in terms of the causal
localization principles, one must show that the lack of reactivity of WS is a
consequence of its intrinsically noncompact localization.

In the context of quantum theory these principles is much more powerful than
their classical counterpart. The concept of \textit{modular localization}
permits to address structural problems of QFT in a completely intrinsic way
which avoids the use of "field-coordinatizations". An illustration of the
power of this relatively new concept is the proof of existence of a certain
class of two-dimensional models starting from the observations that certain
algebraic structures in integrable d=1+1 models can be used to construct
modular localized wedge algebras \cite{AOP}. In the work of Lechner and others
this led to existence proofs for integrable models with nontrivial short
distance behavior together with a wealth of new concepts (see the recent
reviews \cite{Lech} \cite{A-L} and literature cited therein). Even in
renormalized perturbation theory modular localization has become useful in
attempts to replace local gauge theory in Krein space by a positivity
preserving string-local fields in Hilbert space \cite{vector} \cite{beyond}.

In \cite{BGL} it was essential to extract localization properties
\textit{directly} in the form of modular localized subspaces since Weinberg's
group theoretic method of constructing the intertwiners of local field via
group theoretical covariance requirements does not work for WS.

In an unpublished previous note \cite{dark} I tried to address the problem of
a possible connection between WS and dark matter. But the recent gain of
knowledge from modular localization regarding attempts to unite WS with normal
matter under the conceptual roof of AQFT in \cite{LMR}, as well as new
insights coming from perturbative studies of couplings involving string-local
fields \cite{vector} \cite{beyond}, led to a revision of my previous ideas.

In \cite{LMR} it was shown that the attempt to unite normal matter together
with WS in a \textit{nontrivial} way\footnote{Excluding the trivial
possibility of a tensor product of WS with the world of ordinary matter (total
inertness except with respect to classical gravity).} under the conceptual
roof of algebraic QFT (AQFT) leads to an unexpected (suspicious looking) loss
of the so-called "Reeh-Schlieder property" for compact localized observable
algebra. The R-S property states that the set of state vectors obtained by the
application of operators from a \textit{compact} localized subalgebra of local
observables to the vacuum is "total" in the vacuum Hilbert space. The
possibility to manipulate large distance properties of states in the vacuum
sector by applying operators localized in a compact spacetime region
$\mathcal{O}$ to the vacuum is considered to be a universal manifestation of
vacuum polarization.

The R-S property plays an important role in the Doplicher-Haag-Roberts (DHR)
superselection theory \cite{Haag} which leads to the concept of inner
symmetries and Bose/Fermi statistics (absence of parastatistics) and its
absence in the presence of WS asks for further clarification which in the
present work is obtained from a new positivity preserving string-local
perturbation theory (SLFT) for $s\geq1$ fields. It turns out that, different
from the pl case, the pcb requirement on the first order interaction density
$d_{sd}^{int}(L)\leq4$ is the only restriction for the perturbative existence
of a model; string-local (sl) interactions must also fulfill quite restrictive
additional conditions which prevent total delocalization and maintain
renormalizability in higher orders.

The main result of the present paper is that these conditions cannot be
fulfilled in couplings of WS to normal matter. The reason is that the class of
WS remains completely isolated; its fields are not the massless limit of spin
$s\ $massive fields for fixed Pauli-Lubanski invariant $\kappa^{2}%
=m^{2}s(s+1).$

This leaves only the conclusion that, apart from interactions with gravity as
a consequence of the positive energy property and the existence of an
energy-momentum tensor, WS cannot interact with normal matter. In mathematical
terminology: WS tensor-factorizes with normal matter and the Reeh-Schlieder
property holds only in the tensor factor of normal matter.

A world in which the WS matter only reacts with gravity may be hard to accept
from a philosophical viewpoint. But after we got used to chargeless leptons
which only couple to the rest of the world via weak and gravitational
interactions, the step to envisage a form of only gravitationally interacting
kind of matter is not as weird as it looks at first sight.

The paper is organized as follows.

The next section presents a "crash course" on Wigner's theory of positive
energy representations of the Poincar\'{e} group including the explicit
construction of sl WS free fields and their two-point functions.

The third section highlights an important restriction on renormalizable
couplings involving sl fields which is necessary to avoid higher order total delocalization.

In section 4 it is explained why this perturbative restriction is violated for
WS which is the cause of its inertness.

Section 5 addresses the problem of the role of sl localization in the
construction of the correct energy-momentum tensor for higher spin quantum
(i.e. acting in Hilbert space) fields which is the prerequisite for the
Einstein-Hilbert gravitational coupling.

The concluding remarks point at problems arising from the identification of WS
with dark matter.

\section{Matter as we (think we) know it and Wigner's infinite spin "stuff"}

The possible physical manifestations of WS matter can only be understood in
comparison to normal matter. Hence before addressing its peculiarities it is
necessary to recall the localization properties of free massive and finite
helicity zero mass fields.

It is well known that all pl massive free fields can be described in terms of
matrix-valued functions $u(p)$ which intertwine between the
creation/annihilation operators of Wigner particles \cite{Wein}. Their
associated covariant fields are of the form%

\begin{equation}
\psi^{A,\dot{B}}(x)=\frac{1}{\left(  2\pi\right)  ^{3/2}}\int(e^{ipx}%
u^{A,\dot{B}}(p)\cdot a^{\ast}(p)+e^{-ipx}v^{A,\dot{B}}(p)\cdot b(p))\frac
{d^{3}p}{2p_{0}} \label{m}%
\end{equation}
The intertwiners $u(p)$ and their charge-conjugate counterpart $v(p)~$are
rectangular $(2A+1)(2B+1)\otimes(2s+1)$ matrices which intertwine between the
unitary $(2s+1)$-component Wigner representation and the covariant
$(2A+1)(2B+1)~$dimensional spinorial representation labeled by the
semi-integer $A,\dot{B}$ which characterize the finite dimensional
representations of the covering of the Lorentz group $SL(2,C)$. the
$a^{\#}(p),b^{\#}(p)$ refer to the Wigner particle and antiparticle
creation/annihilation operators and the dot denotes the scalar product in the
$2s+1$ dimensional spin space.

For a given physical spin $s~$there are infinitely many spinorial
representation indices of the homogeneous Lorentz group; their range is
restricted by \cite{Wein}%
\begin{equation}
\left\vert A-\dot{B}\right\vert \leq s\leq A+\dot{B},~~m>0 \label{r}%
\end{equation}
For explanatory simplicity we restrict our subsequent presentation to integer
spin $s;$ for half-integer spin there are similar results.

All fields associated with integer spin $s~$representation can be written in
terms of derivatives acting on symmetric tensor potentials ($A=\dot{B}$) of
degree $s$ with lowest short distance dimension $d_{sd}^{s}=s+1.$ For
$s=1~$one obtains the divergenceless ($\partial\cdot A^{P}=0$) Proca vector
potential $A_{\mu}^{P}$ with $d_{sd}=2$, whereas for $s=2$ the result is a
divergence- and trace-less symmetric tensor $g_{\mu\nu}$ with $d_{sd}=3$.

Free fields can also be characterized in terms of their two-point functions
whose Fourier transformation are tensors in momenta instead of intertwiners.
For $s=1$ on obtains
\begin{equation}
\left\langle A^{P}(x)A^{P}(x^{\prime})\right\rangle =\frac{1}{\left(
2\pi\right)  ^{3}}\int e^{-ip(x-x^{\prime})}M_{\mu\nu}^{P}(p)\frac{d^{3}%
p}{2p^{0}},~M_{\mu\nu}^{P}(p)=-g_{\mu\nu}+\frac{p_{\mu}p_{\nu}}{m^{2}}
\label{ma}%
\end{equation}
and for higher spin the $M^{\prime}s$ are symmetric tensors formed from
products of $g_{\mu\nu}$ and of $p^{\prime}s$ (P stands interchangeably for
"Proca" or "point-like")

For $m=0$ and finite integer helicity $h$ the two dimensional $\pm\left\vert
h\right\vert $ helicity representation replaces the $2s+1$ component spin.
Despite this difference, the covariant fields turn out to be of the same form
(\ref{m}), except that (\ref{r}) is now replaced by the more restrictive
relation%
\begin{equation}
\left\vert A-\dot{B}\right\vert =\left\vert h\right\vert ,~~~m=0 \label{0}%
\end{equation}
\textit{which excludes all the previous tensor potentials} but preserves their
field strengths (which are tensors of degree $\left\vert h\right\vert $ and
$d_{sd}=\left\vert h\right\vert +1$ with mixed symmetry properties). This is
well-know in case of $\left\vert h\right\vert =1$ where there exist no
massless pl vector potential $A=1/2=\dot{B}$ who's curl is associated to the
electromagnetic field strength$.$

The absence of pl tensor potentials in (\ref{0}) results from \textit{a clash
between pl spin }$s$\textit{ tensor potentials and} \textit{Hilbert space
positivity}. Gauge theory substitutes the non-existent pl Hilbert space vector
potential by pl potentials in an indefinite Krein space; symmetry unde gauge
tnsfomations prescriptions by which one extracts a physical subtheory from a
Krein space lead to gauge theory. This is also a clash between the classical
Lagrange formalism (positivity has no place in classical physics) and the most
basic Hilbert space positivity on which quantum theory's probability hinges.
It shows the limitation of that parallelism to classical theory called
Lagrangian quantization\footnote{Not to be confused with "second quantization"
which is an unfortunate terminology for a functorial relation between Wigner's
representation theory of particles and the associated quantum free fields
acting in a Wigner-Fock Hilbert space.}.

The problem can be resolved in two ways; either one sacrifices positivity or
one gives the Hilbert space a chance to determine the tightest localization
which is consistent with positivity, which turns out to be localization on
semi-infinite spacelike strings $x+\mathbb{R}_{+}e,~e^{2}=-1$. Beware that
there is no relation between sl fields and string theory. Whereas the change
from pl to sl fields for $s\geq1$ is required in order to uphold Hilbert space
positivity, ST has no conceptual compass, it is the result of a playful spirit
to extend the game of QFT.

The first solution leads to a physically restricted theory in Krein space in
which all gauge dependent fields are physically void. The advantage is only
computational since pl fields are computationally easier (however this does
not apply to the explicit extraction of the physical data with the help of the
BRST ghost formalism which remains involved). The perturbation theory of sl
fields turns out to be more demanding, but as a reward one obtains a full QFT
in which all fields are physical (though, as for pl$\ s<1$ interactions in
Hilbert space, not all operators represent local observables).

sl tensor potentials also exist for massive fields. The sl counterpart of the
pl massive two-point functions (\ref{ma}) turn out to be%
\begin{equation}
M_{\mu\nu}^{s}(p;e.e^{\prime})=-g_{\mu\nu}-\frac{p_{\mu}p_{\nu}e\cdot
e^{\prime}}{(p\cdot e-i\varepsilon)(p\cdot e^{\prime}+i\varepsilon)}%
+\frac{p_{\mu}e_{\nu}}{p\cdot e-i\varepsilon}+\frac{p_{\nu}e_{\mu}^{\prime}%
}{p\cdot e^{\prime}+i\varepsilon} \label{s}%
\end{equation}
Its massless limit is of the same form, except that the momentum $p$ is on the
boundary of the positive lightlike surface $H_{0}^{+}$ of the forward light
cone $H_{m}^{+};$ This has to be taken into account in the Fourier
transformation to $x$-space.

The more complicated form as compared to the simpler $-g_{\mu\nu}$ in the
Feynman gauge setting is the prize to pay for improving the high energy
behavior while \textit{preserving positivity} and securing the existence of a
massless limit. The only way I know which secures positivity is the
intertwiner representation of covariant fields as linear combinations of
Wigner creation/annihilation operators. Lagrangian quantization account for
positivity only for $s<1$ interactions; in all higher spin cases it leads to
indefinite metric which only permits a partial return to positivity in case of
a gauge formalism in Krein spaces. I am not aware that physical (Hilbert
space) $s>1\ $energy-momentum tensors have been constructed in the existing literature.

The best description of the interacting massless theory is to first calculate
the renormalized massive correlation functions and then take their massless
limit. This has the advantage of performing perturbation theory in the simple
Wigner Fock particle space and leaving the reconstruction of the massless
limit (in which this physical description of the Hilbert space in terms of
particle states is lost) to the application of Wightman's reconstruction
theorem to the limiting correlation functions.

Massive theories are simpler from a conceptual viewpoint because the presence
of a mass gap permit to use the tools of scattering theory and the
identification of the Hilbert space with a Wigner Fock space. Whereas it is
plausible that the asymptotic short distance behavior of the Hilbert space
setting is correctly accounted for in terms of the asymptotic freedom
properties of gauge theories, the problems related to long-distance properties
as confinement remain outside the physical range of the gauge setting.

There is a very efficient way to derive the relation between the pl Proca
potential and its sl counterpart. Integrating the latter along the space-like
direction $e,$ one obtains a sl scalar field $\phi(x,e)$
\begin{align}
\phi(x,e)  &  :=\int_{0}^{\infty}d\lambda e^{\mu}A_{\mu}^{P}(x+\lambda
e)=\frac{1}{(2\pi)^{3/2}}\int(e^{ipx}u(p,e)\cdot a(p)+h.c.)\frac{d^{3}%
p}{2p_{0}}\label{vi}\\
u(p,e)  &  :=u(p)\cdot e\frac{1}{ip\cdot e},~~M^{\phi,\phi}=\frac{1}{m^{2}%
}-\frac{e\cdot e^{\prime}}{(p\cdot e-i\varepsilon)~(p\cdot e+i\varepsilon)}%
\end{align}
where the inner product in the first line refers to the 3-dim. spin space and
the denominator is simply the Fourier transform of the Heavyside function. The
$\phi$ two-point function can either be computed from carrying out the line
integrals on the Proca two-point function or by using the intertwiner. The sl
vector is defined as
\begin{equation}
~A_{\mu}(x,e):=\int_{0}^{\infty}e^{\nu}F_{\mu\nu}(x+\lambda e)ds,\,\text{\ }%
F_{\mu\nu}:=\partial_{\mu}A_{\nu}^{P}-\partial_{\nu}A_{\mu}^{P} \label{field}%
\end{equation}
which leads to the two-point function (\ref{s}).

The three fields turn out to be linearly related%
\begin{equation}
A_{\mu}(x,e)=A_{\mu}^{p}(x)+\partial_{\mu}\phi(x,e) \label{rel}%
\end{equation}
which either can be derived from the previous definition or by defining the
three fields in terms of their intertwiners in which case (\ref{rel}) is a
numerical relation between the numerical intertwiner functions. A similar
looking relation in which the sl vector potential is replaced by the
Gupta-Bleuler pl gauge potential and the $\phi$ by the negative metric
St\"{u}ckelberg field whose two-point function is the negative of a scalar pl
field \cite{Ruegg}. In the presence of interactions one needs additional ghost
degrees of freedom. In a somewhat metaphoric terminology one may say that the
positivity preserving sl field theory (SLFT) is the result of applying Okhams
razor to Gauge Theory (GT).

We will refer to fields which mediate between pl potentials and their sl
counterparts as \textit{escorts}. When passing from positivity preserving
$m=0$ potentials to massive Proca potentials it is not enough to "turn on the
mass" but one also needs the intervention of the escorts. They play exactly
the role which is erroneously attributed to the Higgs field ("fattening the photons").

By changing the $\lambda$-measure $d\lambda\rightarrow\kappa(\lambda)d\lambda$
one can improve the short distance behavior and get arbitrarily close to
$d_{sd}=0,$ but this will be of no avail\footnote{In particular the lowering
of $d_{sd}$ is of no help for $s<1$ fields. There is no Elko trick which
improves the short distance properties of $s=1/2$ fields. The proposal in
(\cite{Elko}, formula ()) shows a total misunderstanding of what QFT is
about.} since this would destroy the linear relation (\ref{rel}) which permits
to use the lower $d_{sd~}$ to achieve renormalizability (the $L,V_{\mu}$
condition in the next section).$~$

The relation (\ref{rel}) looks like a gauge transformation; indeed the
extension to interactions with matter fields suggests a formal connection
between pl $\psi(x)$ and is sl counterpart which has the expected exponential
form $\psi(x,e)=\psi(x)\exp ig\phi(x,e$) But in contrast to gauge theory these
relations intertwine between sl fields and their more singular pl siblings
within the same sl relative localization class; in fact this formula, after
making it precise in terms of normal products, may be seen as the definition
of an $e$-independent $d_{sd}=\infty$ singular pointlike counterpart of a
polynomially bounded sl field. The singular nature of the pl "field
coordinatization"~leads to the typical with perturbative order increasing
number of counter-term parameters whereas in the sl coordinatization the
number of parameters remains finite, just as in renormalizable pl $s<1$ interactions.

This construction can be extended to all integer spin fields \cite{M-O} (and
with appropriate modifications also to Fermi-fields). The divergence-free
Proca potential is replaced by divergence- and trace-free symmetric tensor
potentials $A_{\mu_{1},...\mu_{s}}$ of tensor rank $s$. Iterated integration
along a space-like direction $e$ leads to $s$ sl $\phi~$tensor fields of lower
rank
\begin{equation}
\phi_{\mu_{1}..\mu_{k}}(x,e)=\int d\lambda_{1}..d\lambda_{s-k}e^{\nu_{1}%
}..e^{\nu_{s-k}}A_{\nu_{1},...\nu_{s-k},\mu_{1},...\mu_{k}}(x+\lambda
_{1}e+..\lambda_{s-k}e) \label{multi}%
\end{equation}
Again one can construct the $e$- dependent intertwiners from these relations.
The extension of (\ref{rel}) spin $s$ relates the pl tensor-potential $A^{P}$
to its sl counterpart and the symmetrized contributions from the derivatives
of the sl tensor escorts$~\phi...(x,e)$ of $A....(x,e)$ (\ref{multi})%
\begin{align}
A_{\mu_{1}},.._{\mu_{s}}(x,e)  &  =A_{\mu_{1}}^{P},.._{\mu_{s}}(x)+sym\sum
_{k=1}^{s}\partial_{\mu_{1},..}\partial_{\mu_{k}}\phi_{\mu_{k}+1,..\mu_{s}%
}\label{ten}\\
g_{\mu\nu}(x,e)  &  =g_{\mu\nu}^{P}(x,e)+sym\ \partial_{\mu}\phi_{\nu
}+\partial_{\mu}\partial_{\nu}\phi
\end{align}
where the second line is the special case of the connection between the trace-
and divergence-less $s=2$ pl symmetric tensor and its sl counterpart including
the two sl $s<2$ escorts $\phi_{\mu}$ and $\phi.$

The appearance of these lower spin $\phi$-escorts is characteristic for the
change of massive pl fields into their sl siblings acting in a Hilbert space.
They are important new fields which depend on the same degrees of freedom (the
Wigner creation/annihilation operators) as the other two operators.

For $s=1~$the scalar escort field $\phi$ may be seen as the QFT analog of the
bosonic Cooper pairs which are the result of a reorganization of the condensed
matter degrees of freedom in the superconducting phase. Without the formation
of Cooper pairs from existing condensed matter degrees of freedom it is not
possible to convert the long-range classical vector potentials into its short
range counterparts within the superconductor (as anticipated by London).

The relation between long range massless and short range pl Proca potentials
requires the presence of the $\phi;~$in fact it is not possible to formulate
massive QED as a renormalizable theory in Hilbert space without the presence
of these scalar escorts, and where there is need of additional $H$ fields, as
for massive self-interacting vector mesons, it is for entirely different
physical reasons than spontaneous symmetry breaking.

The reason why additional degrees of freedom in the form of $H$\textit{-fields
are indispensable in the case of self-interacting massive vectormesons} is
quite deep but bears no relation to physical spontaneous symmetry breaking.

The lower spin escort fields in (\ref{ten}) have no massless limit, but
together with the Proca tensor potentials their presence is necessary for the
construction of the massive sl potential; all these fields are relatively
local and act in the same Wigner-Fock Hilbert space. Only the correlation
functions of the degree $s$ sl tensor potential possess a massless limit.

The sl $A...$ in (\ref{ten}) is related to the pl field strength%
\begin{equation}
\mathcal{F}_{\mu_{1}..\mu_{s},\nu_{1}..\nu_{s}}=\underset{\mu,\nu}%
{~as}\left\{  \partial_{\mu_{1}}..\partial_{\mu_{s}}A_{\nu_{1},..\nu_{s}%
}\right\}  \label{F}%
\end{equation}
where the $as$ imposes antisymmetry between the $\mu-\nu$ pairs. The
antisymmetrization effects the $d_{sd}$ of the pl tensor potential is the same
as that of field strengths namely $d_{sd}=s+1$. As in the previous case of the
vector potential (\ref{field}), the sl tensor potentials (\ref{ten}) can be
obtained in terms of$~$iterated integrations along $e$ starting from the field
strength. The field strength tensor is the lowest rank pl field which has a pl
massless limit\footnote{For $s=2$ this tensor has the same mixed symmetry
property as the linearized Riemann tensor whereas the symmetric second rank
tensor deserves to be denoted as $g_{\mu\nu}.$}. With appropriate changes
these results have analogs for semi-integral spin.

Before passing to the sl fields of the WS class it may be helpful to collect
those properties which turn out to be important for higher spin sl fields.

\begin{itemize}
\item Whereas pl massive tensor potentials have short distance dimension
$d_{sd}=s+1,$ their sl counterparts have $d_{sd}=1$ independent of spin. Hence
there are always first order sl interaction densities within the
power-counting limit $d_{sd}^{int}\leq4,$ but whether they can be used in a
consistent perturbative renormalization setting is another story.

\item Sl tensor potentials are smooth $m\rightarrow0$ limits of their sl
massive counterpart; they inherit in particular the $d_{sd}=1$ from their
massive counterpart. The lowest pl fields in the same representation class
are$~$field strengths (tensors of rank $2s$ and $d_{sd}=s+1~$with mixed
symmetry properties).

\item The pl $d_{sd}^{K}=1~$zero mass vector potentials $A_{\mu}^{K}$ of local
gauge theory act in an indefinite metric Krein space. The physical price for
resolving the clash between point-like localization and Hilbert space
positivity is the is the loss of both the positivity and the correct physical
localization whose validity is restricted to gauge invariant observables. What
makes gauge theory useful for particle theory (the Standard Model) is the fact
that the perturbative unitary on-shell S-operator is gauge invariant. The
absence of local observables excludes the use of gauge theory in WS models.
\end{itemize}

Whereas the two-point functions of pl massive free fields are polynomial in
$p,$ their sl counterparts have a rational $p$-dependence (\ref{s})
(\ref{vi}). The family of all WS interwiners for a given Pauli-Lubanski
invariant $\kappa~$has been computed in \cite{MSY}, their two point-functions
are \textit{transcendental functions} of $p,e$ which are boundary values of
from $Im(e)\in V^{+}.$

A particular simple WS intertwiner with optimal small and large momentum space
behavior (corresponding to the minimal choice for sl massless $s\geq1$) has
been given in terms of an exponential function in \cite{Koe}%
\begin{equation}
u(p,e)(k)=expi\frac{\vec{k}(\vec{e}-\frac{p_{-}}{e_{-}}\vec{p})-\kappa}{p\cdot
e} \label{Ko}%
\end{equation}
here $k$ is a two-component vector of length $\kappa$; the~Hilbert space on
which Wigner's little group $E(2)~$acts consists of square integrable
functions $L^{2}(k,d\mu(k)=\delta(k^{2}-\kappa^{2})dk)$ on a circle of radius
$\kappa$. The vector arrow on $e$ and $p~$refer to the projection into the
$1$-$2$ plane, and the and $e_{-},p_{-}$ refer to the difference between the
third and zeroth component. The most general solution of the intertwiner
relation differs from this special one by a function $F(p\cdot e)$ which is
the boundary value of a function which is analytic in the upper half-plane
\cite{MSY}.

The two-point function is clearly a $J_{0}~$Bessel function. The calculation
in \cite{Koe} was done in a special system. Writing its argument in a
covariant form one obtains \footnote{I am indepted to Henning Rehren for
showing me the covariantization of K\"{o}hler's result.}%

\begin{align}
M^{WS}(p,e)  &  \sim J_{0}(\kappa\left\vert w(p,e)\right\vert )exp-i\kappa
(\frac{1}{p\cdot e-i\varepsilon}-\frac{1}{p\cdot e^{\prime}+i\varepsilon
})\label{two}\\
with~w^{2}(p,e)  &  =-(\frac{e}{e\cdot p-i\varepsilon}-\frac{e^{\prime}%
}{e^{\prime}\cdot p+i\varepsilon})^{2}\nonumber
\end{align}
The exponential factor compensates the singularity of $J_{0}~$at $e\cdot
p=0.~$Note that the Pauli-Lubanski invariant $\kappa$ has the dimension of a
mass so that the argument of the two-point function of the sl field has the
correct engineering dimension $d_{en}=1$ of a quantum field.

The main purpose of this calculation is to convince the reader that there are
explicitly known transcendental WS intertwiner and two-point functions whose
associated propagators have a well behaved ultraviolet and infrared behavior.
As already mentioned, the physical reason why these fields are nevertheless
excluded from appearing in interaction densities is that higher orders lead to
a complete delocalization; this will be explained in the next section.

\section{The problem of maintaining higher order sl localization}

It is well known that the only restriction for pl interaction densities is the
power-counting inequality $d_{sd}^{int}\leq4.$ Since the minimal short
distance dimension of pl spin $s$~fields\footnote{Fields (without the added
specification "gauge") are always acting in Hilbert space.} is $s+1,~$there
are no pl renormalizable interactions involving $s\geq1~$fields. Sl free
fields on the other hand have an $s$-independent short distance behavior
$d_{sd}=1,$ so that one always can find polynomials of maximal degree 4 which
represent interaction densities within the power-counting limitation.

Point- and sl fields represent two different descriptions of the same spin $s$
quantum matter, just like two different coordinatization in differential
geometry. For $s\geq1$ the use of the pl coordinatization in interaction
densities becomes too singular; the breakdown of the power-counting bound
$d_{sd}^{int}\leq4$ leads to singular interacting fields (unbounded increase
of $d_{sd}$ with perturbative order). The bad aspect of such a singular (non
Wightman) behavior is not primarily the polynomial unboundedness in momentum
space, but rather the fact that the perturbative counterterm formalism leads
to an ever increasing number of undetermined counterterm parameters which
destroys the predictive power.

There are two explanations for the cause of this situation, either the
interaction density is incompatible with the principles of QFT or the model is
consistent but the pl coordinatization is too singular for the application of
the rules of renormalized perturbation theory. In the latter case the use of
sl field coordinatization may lead to reduction of the short distance
singularity within $d_{sd}^{int}\leq4$ and in this way save the model. If this
fails one falls back to square one, this time without remedy.

Fields which even in their sl coordinatization do not lead to renormalizable
couplings cannot be used for defining interaction densities of renormalizable
model will be referred to as \textit{inert}. The main claim of the present
work is that WS fields fall into this category whereas for higher finite spin
fields $s\geq2\ $the question whether they are "reactive" or remain inert
remains unsettled (see next section).

Formally renormalizable sl interaction densities within the power-counting
bound come with a physical hitch. Unless they fulfill an additional
requirement it is not possible to maintain the sl localization in higher
orders. In that case the result will be a \textit{complete delocalization} and
hence the principles of QFT exclude such interactions. In the next section it
will be shown that the presence of WS fields in a interaction density leads to
such a situation.

On the other hand free WS fields fulfill all general localization- and
stability- requirements (energy-positivity) of QFT and consequently cannot be
excluded as being unphysical \cite{LMR}. This justifies the terminology "inert
matter" used in the next section. The remainder of this section will address
the problem of upholding sl localization in higher orders which will be taken
as the defining property of "reactive (or dynamic) quantum matter".

As a simple nontrivial illustration of this additional delocalization
preventing requirement we start with a sl interaction density $L\ $of massive
QED%
\begin{equation}
L=A_{\mu}(x,e)j^{\mu}(x) \label{Q}%
\end{equation}
Here $A_{\mu}(x,e)$ is a massive sl vector potential (\ref{s}) and $j^{\mu}$
is the conserved current of a massive complex scalar field. These fields act
in a Wigner-Fock Hilbert space of the corresponding Wigner particles, and
since $d_{sd}(A_{\mu})=1~$and hence the short distance dimension $d_{sd}%
^{int}(L)=4,\ L$ in (\ref{Q}) stays within the power-counting bound
$d_{sd}^{int}=4$ of renormalizability. The sl $L$ is related to its $d_{sd}=5$
pl counterpart $L^{P}$ as%

\begin{align}
&  L^{P}=A^{P}\cdot j=L-\partial^{\mu}V_{\mu},~~~V_{\mu}(x,e):=\partial_{\mu
}\phi j^{\mu}\label{scalar}\\
&  \int L^{P}d^{4}x=\int Ld^{4}x,~i.e.~S^{(1)}=S_{P}^{(1)}=S_{S}^{(1)}
\label{S}%
\end{align}
where the second line follows since in the presence of a mass gap the
divergence of $V_{\mu}$ does not contribute to the adiabatic limit which
represents the first order S-matrix. In other words one splits the
$d_{sd}=5\ $pl density into its sl $d_{sd}=4\ $counterpart and a $d_{sd}%
=5\ $divergence term which can be disposed of in the adiabatic (on-shell)
S-matrix limit.

In this way one solves two problems in one stroke, on the one hand one
expresses the (first order) S-matrix in terms of a $d_{sd}=4$ interaction
density, and at the same time the $e$-dependence disappears in the first order
on-shell S-matrix. The linear relation between $L$ and its pl counterpart
$L^{P}$ is a consequence of the linear relation between the massive pl Proca
potential and its sl $d_{sd}=1\ $counterpart (\ref{rel}). The lowering to
$d_{sd}<1$ by using instead of $d\lambda$ another measure $\mu(\lambda
)d\lambda$ would be possible, but this would destroy the linear relation
(\ref{scalar}) and as a result also the relation (\ref{S}) which is the basis
of the $e$-independence of $S$\footnote{Recent suggestions \cite{Elko} that
the mere lowering of $d_{sd}=3/2\ $to $1$ for $s=1/2$ (probably to lower the
power-counting bound of the 4-Fermi interaction) reveal a misunderstanding of
QFT.}$.$

For the following it is convenient to formulate the $e$-independence in terms
of a differential calculus on the $d=1+2$ dimensional directional de Sitter
space. The differential form of the relation (\ref{rel}) reads%
\begin{align}
d_{e}A_{\mu}  &  =\partial_{\mu}u,~~u=d_{e}\phi\label{1}\\
d_{e}(L-\partial^{\mu}V_{\mu})  &  =d_{e}L-\partial^{\mu}Q_{\mu}=0,~Q_{\mu
}:=d_{e}V_{\mu} \label{first}%
\end{align}
Hence $A$, $\phi,L,V$ are $d_{sd}=1$ zero-forms whereas $u,Q$ are exact
$d_{sd}=1$ one-forms; together with the exact two-form $\hat{u}\ $(\ref{vi})
they exhaust the linear with $A_{\mu}^{P}$ linear related relatively local
$d_{sd}=1$ forms.

We will refer to the relation (\ref{first}), which expresses the
$e$-independence in terms of a closed zero form, as the "$L,V$ (or $L,Q$)
relation" (\ref{first}). It is a necessary condition for the $e$-independence
of $S.$ Its extension to vacuum expectation values of fields requires that
they depend only on those $e^{\prime}s$ of the fields and not on the
$e^{\prime}s\ $of inner propagators in Feynman diagrams which contribute to
these correlations.

As the independence of $S$ from gauge-fixing parameters in gauge theory, this
$e$-independence in the Hilbert space setting results from cancellations
between different contributions in the same order; but different from
unphysical gauge dependent correlation functions, correlations of
charge-carrying sl fields are expectation values of physical fields in an
extended Wightman setting (endpoint $x$- \textit{and} directional $e$-smearing).

In order to secure the $e$ independence in higher orders we must extend the
$L,Q_{\mu}~$relation (\ref{first}) to higher order time-ordered products. The
second order $L,Q_{\mu}$ pair requirement\footnote{The $Q_{\mu}$ formalism is
somewhat simpler than its $V_{\mu}$ couterpart. For massive QED and couplings
of massive vector mesons to Hermitian matter ("Hermitian QED", the Higgs
abelian model) it is easy to see their equivalence.} reads
\begin{equation}
(d_{e}+d_{e^{\prime}})TLL^{\prime}-\partial^{\mu}TQ_{\mu}L^{\prime}%
-\partial^{\mu^{\prime}}TLQ_{\mu}^{\prime}=0 \label{2}%
\end{equation}
If it where not for the distributional singularities of $T$-products at
coalescent points, this would follow from (\ref{first}). For the second order
S-matrix we only need the one particle contraction component ("tree approximation").

For massive spinor QED the relation is fulfilled in term of the standard free
field propagator. The more singular scalar QED contains $d_{sd}=2$ derivatives
$\partial\varphi$ which according to the minimal scaling rules of the
divergence and regularization free Epstein-Glaser renormalization theory lead
to a delta counterterm%
\begin{equation}
\left\langle T\partial_{\mu}\varphi^{\ast}\partial_{\nu}^{\prime}%
\varphi^{\prime}\right\rangle =\partial_{\mu}\partial_{\nu}^{\prime
}\left\langle T\varphi^{\ast}\varphi^{\prime}\right\rangle +cg_{\mu\nu}%
\delta(x-x^{\prime})
\end{equation}
The imposition of the relation (\ref{2}) fixes the parameter $c$ with the
expected result of an induced second order term$\ g^{2}g\delta(x-x^{\prime
})A_{\mu}A^{\mu}\varphi^{\ast}\varphi.$

Note that no arguments of classical gauge theory (as the replacement
$\partial\rightarrow D=\partial+igA$) has been used; the result is solely a
consequence of the causal localization principles and Hilbert space positivity.

There are some interesting foundational aspects of this otherwise trivial
calculation. The independent fluctuation in $e$ and $e^{\prime}$ do not allow
to set $e=e^{\prime}$ in off-shell correlations (\ref{s}); the different
$i\varepsilon$ prescriptions for $e$ and $e^{\prime\ }$in the off-shell
propagator prevent this; apart from the fact that unlike the dependence in
$x~$the\ two-point function in $e$ does not depend on the difference
$e-e^{\prime}~$the less singular behavior at coinciding direction is similar
to those at coalescent points. The remedy in both cases is to use Wick-products.

The on-shell $e$-independence corresponds to the second order gauge invariance
of the scattering amplitude; individual contributions are generally
$e$-dependent and upon setting $e=e^{\prime}$ lead to infinite fluctuations.

The "magic" of $L,V_{\mu}$ pairs with (\ref{first}) is that on the one hand
they permit to use the lower short distance dimension of sl fields (and in
this way lower the power-counting bound of renormalizability) and on the other
hand they also guaranty the $e$-independence of the S-matrix since the
derivative contributions in (\ref{2}) disappear in the adiabatic S-matrix
limit
\begin{equation}
(d_{e}+d_{e^{\prime}})S^{(2)}=0,\ \ S^{(2)}\sim\int TLL^{\prime}%
\end{equation}

The extension of the adiabatic limit of the Bogoliubov $S(g)$ operator
functionals to quantum fields leads to correlation functions of interacting sl
fields. As the S-matrix is independent of the $e^{\prime}s$ of the inner
propagators (after summing over sufficiently many contributions in a fixed
perturbative order), the correlation functions of interacting fields only
depend on the $e^{\prime}s$ of those fields.

A new phenomenon is that the higher order interactions spread the
$e$-dependence also to those fields which entered the first order interaction
density as $s<0~$pl fields\footnote{The perturbation theory of interacting
string-local fields is still in its beginnings. A mathematically rigorous
presentation will be the subject of forthcoming work by Jens Mund..}. In fact
the interacting matter fields in the new sl setting of renormalization theory
are sl in a stronger sense than the vector potentials which remain linearly
related with their pl field strengths.

In the limit of massless sl vector mesons the correlation functions change
their physical properties define a very different theory; the particle setting
in a Wigner-Fock Hilbert space disappears and the strings of the
charge-carrying fields become more "stiff" and cause a spontaneously breaking
of Lorentz invariance in charged sectors \cite{Froe}. Little is known about
the spacetime aspects of these physical changes (particles$\rightarrow
$infraparticles) apart from prescription in momentum space for
photon-inclusive cross sections \cite{YFS}.

The $L,V_{\mu}$ (or $L,Q_{\mu}$) pair property (\ref{1}) is a necessary
condition for maintaining sl localization; it permits to sail between Scilla
of nonrenormalizability and the Charybdis of total delocalization.
Heuristically speaking it provides a compensatory mechanism between
contributions to the same order which prevents the total delocalization
resulting from the integration over $x$ in inner strings $x+\mathbb{R}_{+}e$
in individual Feynman diagrams$.\ $The main point of the present work is the
argument that in the presence of WS fields in $L$ it is not possible to
fulfill the $L,V_{\mu}$ condition so that WS matter can only exist in the
interaction-free form. We will refer to such matter as \textit{inert }(next section).

There is another important physical aspect of the $L,V_{\mu}$ pair property in
which the escort field $\phi$ plays an essential physical role even though it
does not add new degrees of freedom. Heuristically speaking the transition
from long range massless sl vector potentials to their short range massive
counterpart is not possible without the appearance of the $\phi$ escort. The
escort appears explicitly in the pair condition; in some models they already
appear in the first order interaction density (see \ref{ab} below).

This is somewhat reminiscent of the presence of the bosonic Cooper pairs in
the BCS description of superconductivity; without their presence it is not
possible to convert long range classical vector potentials into their short
ranged counterparts inside the superconductor. As the $\phi$ in massive QED
they are not the result of additional degrees of freedom, they rather arise
from rearrangements of existing condensed matter degrees of freedom in the low
temperature phase.

The QFT analog of the BCS or the Anderson screening mechanism is the screened
"Maxwell charge" \cite{vector} i.e.
\begin{align}
j_{\mu}  &  :=\partial^{\nu}F_{\mu\nu},\ Q_{scr}=\int j_{0}(x)d^{3}x=0\\
\partial^{\mu}j_{\mu}  &  =0,\text{ }Q_{SSB}=\int j_{0}(x)d^{3}x=\infty,\text{
}long\ dist.divergence
\end{align}
The screening property (first line) \textit{only depends on the massive field
strength and not on the kind of matter to which it couples} (which may be
complex or Hermitian matter). This includes non-interacting massive vector
mesons for which $j_{\mu}\sim A_{\mu}^{P}.$ Spontaneous symmetry breaking on
the other hand reveals itself in form of a conserved current whose charge
diverge instead of being zero (second line).

Renormalizable models are generally uniquely specified in terms of their field
content. In the above case of massive scalar QED the form of the first order
sl coupling is uniquely fixed by the $L,Q_{\mu}$ pair condition (the
preservation of sl localization). The second order pair condition is a
normalization requirement which \textit{induces} the $A\cdot A\left\vert
\varphi\right\vert ^{2}~$term.

This induction which in classical QED results from the fibre-bundle structure
$\partial_{\mu}\rightarrow D_{\mu}=\partial_{\mu}-igA_{\mu}$ is in (positivity
preserving) QFT a \textit{structural consequence of the causal localization
principle}. Gauge theory hides this important fact by preserving the classical
fibre bundle interpretation at the price of indefinite metric (which does not
only violate the probability interpretation but also denaturalizes the
physical localization of QFT). The role of gauge symmetry and gauge invariance
is to recover part of the lost physical properties for a more restricted gauge
invariant subtheory. The induction of the quadratic $A\cdot A\left\vert
\varphi\right\vert ^{2}$ is a result of the imposed invariance under "gauge
symmetry" which (apart from the fact that this is not a physical symmetry) is
is equivalent to the classical fibre bundle requirement.

The new SLFT setting retains all physical properties by replacing the mute
global gauge fixing parameters by individually fluctuating local space-like
string directions. In this way all the unphysical "dead wood" of indefinite
metric St\"{u}ckelberg fields and ghosts will not be allowed to enter in the
first place. The idea that quantum fields should follow the pl localization of
classical field theory is too restrictive for constructing interacting higher
spin quantum fields.

The sl localization preserving $L,V_{\mu}$ induction becomes much richer if
the interaction of $A_{\mu}$ with complex matter is replaced by Hermitian
matter (the abelian Higgs model). The application of the $L,V$ requirement to
the coupling of a massive vector meson to a Hermitian $H$ field proceeds as
follows. The pl interaction with the lowest short distance dimension$\ d_{sd}%
^{P,int}=5$ is $L^{P}=mA^{P}\cdot A^{P}H.$ Converting it into a sl $L,V_{\mu
}~$pair, one obtains (easy to check by the use of the free Klein-Gordon
equation for $H$ and relation (\ref{rel})):
\begin{align}
L  &  =m\left\{  A\cdot(AH+\phi\overleftrightarrow{\partial}H)-\frac{m_{H}%
^{2}}{2}\phi^{2}H\right\}  ,~V_{\mu}=m\left\{  A_{\mu}\phi H+\frac{1}{2}%
\phi^{2}\overleftrightarrow{\partial}_{\mu}H\right\} \label{ab}\\
&  L-\partial V=L^{P}=mA^{P}\cdot A^{P}H,\text{ ~}d_{e}(L-\partial
V)=0\nonumber
\end{align}
In this case the on-shell $e$-independence requirement (\ref{2}) in second and
third order tree approximation leads to a much richer collection of induced
terms than that of scalar massive QED \cite{vector} (for a gauge-theoretic
derivation of the induction see \cite{Scharf} section 4.1).

Whereas in the massive scalar QED model this requirement induces only the
$A\cdot A\varphi^{\ast}\varphi$ term, the induction in case of an interaction
with a Hermitian field leads besides the expected $A\cdot AH^{2},A\cdot
A\phi^{2}$ terms (which as in scalar QED can be absorbed into a changed
time-ordered product) also to second order induced $H^{4},\phi^{4},H^{2}%
\phi^{2}$ terms (from second order $A$-$A$ contractions in (\ref{2}) as well
as to an additional first order $H^{3}$ term \cite{vector} \cite{beyond}. The
coupling strengths of these second order\footnote{In higher ($4^{th}$) order
one also expects the appearance of new counterterm parameters as known from
point-local interactions.} induced terms are fixed in terms of the 3 physical
parameters of the elementary model-defining $A_{\mu},H$ fields, namely the
coupling strength and (ratios of) the two masses $m,m_{H}.$

The result corresponds to the terms induced by gauge invariance of the
S-matrix in \cite{Scharf}. It is also the same as that of the formal
calculation based on the SSB Higgs mechanism, except that in that case one
postulates a Mexican hat potential instead of inducing it from gauge
invariance or from causal localization in a positivity preserving sl setting.
As soon as vector potentials enter one has to follow the rules of either GT or SLFT.

QFT is a foundational quantum theory in which all physical properties of a
model are intrinsic; for the physical interpretation of its content one is not
forced to rely on prescriptions. In many cases it is the field content alone
which determines the form of the interaction density. For the case at hand the
$A,H$ field content and the $L,Q_{\mu}$ renormalization requirement fix the
first and second order interaction density including the $H$ self-interaction
which is erroneously attributed to a SSB.

The shift in field space on a Mexican hat potential is a useful trick whenever
the model permits a SSB i.e. whenever a conserved current leads to a (long
distance) divergent charge $Q_{ssb}=\infty~$(the definition of SSB). This is
the case as long as the Mexican hat potential is not coupled to a vector
potential (or any other $s\geq1$ potential). The $A_{\mu}$ coupling changes
this since the only conserved current of interactions of abelian massive
vector mesons with \textit{any} matter (massive QED, $H$-matter) is the
identically conserved Maxwell current of a massive $F_{\mu\nu}$ field which
always leads to a screened charge $Q=0$. In other words the "Mexican hat +
shift in field space trick" leads to a SSB $L$ only if the field content
permits a SSB ($\partial j=0,Q=\infty$). In that case it is useful to
determine an $L_{SSB}$ from a $L_{SYM}$ (maintaining the symmetry in the
quadrilinear terms while causing a current conservation preserving change in
the lower degree contributions)$.$

To refer to classical gauge symmetry as a "local symmetry" makes perfect
physical sense which is however lost in QFT where gauge invariance is a formal
device to extract a physical subtheory (local observables, S-matrix) from an
unphysical indefinite metric setting. The full QFT in which \textit{all}
fields are physical can be obtained by fighting the increase $d_{sd}=s+1~$of
short distance singularities by using the only \textit{physical} resource of
lowering $d_{sd}$ namely passing from pl Wigner fields. The correct sl fields
are those which lead to the $L,Q_{\mu}$ renormalization theory. Neither the
gauge prescription nor the SLFT setting offer a physical arena for SSB.

One reason why in GT the BRST invariance of the on-shell S-matrix which leads
to the correct induced $H$ self-interactions is easily confused with the
off-shell Mexican hat prescription is that in functional Feynman graph
representation it is difficult to distinguish between relations which only
hold on-shell from off-shell relations. For this reason it was important to
use the causal gauge invariant (CGI) operator (Epstein-Glaser) formulation of
the BRST formalism \cite{Scharf} \cite{BDSV}.

There is no problem to extend the construction of the $L,Q$ pair to
self-interacting massive vector mesons $A_{\mu}$ and calculate the second
order induced terms. \textit{One finds that there is an uncompensated }%
$d_{sd}=5$\textit{ induced term}. Such a nonrenormalizable second order
contribution is deadly if there would be no possibility to extend the field
content of the model in such a way that the interaction of $A_{\mu}$ with the
new field leads to a compensating second order $d_{sd}=5$ contribution. The
new field should have a lower spin (in order not to worsen the short distance
situation) and the same Hermiticity property as $A_{\mu}$ i.e. it must be a
$H$-field.

The compensation against another second order induced term from a first order
$AAH~$interaction works and converts the extended model into a renormalizable
sl QFT \cite{vector} \cite{beyond} (or \cite{Scharf} in gauge theoretical
setting in Krein space). It attributes a fundamental role to the $H$ coupling
which is consistent with the principles of QFT\footnote{Note that $d_{sd}=5$
contributions which have to be compensated do not occur in SSB models.}. This
compensating field is the Higgs field and the compensation is its raison d'\^{e}tre.

This higher order compensation is a new phenomenon of $s\geq1$ sl interactions
which has no analog in $s<1~$pl interactions. Both the $L,Q_{\mu}$ pair
condition as well as the higher order compensation mechanism are the
prerequisites for the concepts of reactive and inert $s>1~$matter in the next section.

\section{Reactive and inert fields for $s~\geq1$}

There are good reasons to believe that problems of lack of convergence of its
power series in coupling parameters are related to the singular nature of the
quantum fields (which are the objects which one expands). This view is
supported by recent existence proofs for two-dimensional integrable models.
These proofs are based on top-to-bottom constructions i.e. they start from
on-shell objects as the S-matrix and pass to operator algebras with the help
of \textit{modular localization} \cite{A-L} and avoid the use of objects as
fields (whose singular behavior in the presence of interactions is caused by
vacuum polarization clouds which result from their application to the vacuum).

Leaving the problem of the possible cause of divergence of perturbative series
aside we will follow the standard parlance of particle theory and identify the
existence and physical properties of a model with those of its renormalized
perturbation theory

The fact that positivity-preserving pl renormalizable models exist only for
$s<1$ interactions and that the use of $s\geq1$ fields in renormalizable
interaction densities requires to use them in their sl form leads however to
peculiarities which play an important role in the division into reactive and
inert fields. Here inert fields are fields which do not admit any
renormalizable couplings with themselves nor with other fields. They only
exist as free fields and since particle counters can only register particles
which interact they remain invisible.

Their "darkness" does however not impede their coupling to gravity and their
ability to make their presence felt in the form of gravitational
back-reactions through Einstein-Hilbert couplings of the energy momentum
tensor. The claim that all positive energy matter possesses an energy-momentum
tensor will be addressed in the next section.

For the division into reactive and inert matter the sl localization preserving
$L,Q_{\mu}$ property and the higher order renormalizability-preserving
$d_{sd}\leq5~$compensation mechanism are indispensable. Without the existence
of the compensatory $H$ field, massive vector mesons could not be
"self-reactive". It is immediately clear that $s<1$ fields are reactive since
they admit renormalizable pl interactions.

It is not possible to decide whether a $s\geq1$ field is reactive or inert
without looking into the details of sl renormalization theory. The $L,Q_{\mu}$
pair condition with $d_{sd}(L)\leq4$ encodes $s$-dependent renormalizability
violating short distance contributions into the $Q_{\mu}~$which becomes
disposed in the adiabatic limit. From second and higher order tree
approximations one learns that there are induced terms which lead to a
modification of $L.$ For $s>1$ these induced terms may lead to second order
modification of the original $L$ (see remarks after (\ref{ab})).

In case the $d_{sd}$ of these induced terms is larger than 4 on has to look
for compensating extensions by enlarging the model's field content. Since the
coupling to spin$\ \geq$ $s$ fields worsens the renormalizability properties
only additional couplings with fields of lower spin can save the model.

\begin{definition}
A massive spin $s\geq1$ field is called reactive if it possesses sl
renormalizable interactions with lower spin fields. Fields which are not
reactive will be referred to as inert.
\end{definition}

We have seen that $s=1~$fields are not only reactive with respect to $s<1$
fields but they are also self-reactive provided that their self-interaction is
accompanied by short distance compensating $A$-$H$ interactions. On the other
hand WS matter is inert since massless sl WS fields cannot be obtained as
limits of massive fields. Whereas for each finite $s>1$ the problem of whether
the pair condition obeying couplings to lower spin fields are possible remains
open, it is easy to see that the pair condition cannot be fulfilled in the
presence of WS matter since the $d_{e}\ $its sl field $\Psi^{WS}%
(x,e)~$(\ref{two}) is not of the form of a divergence $\partial^{\mu}\phi
_{\mu}^{WS}$ which is the prerequisite for the existence of a $L,Q_{\mu}%
~$pair. 

The difficulties of finding reactive spin $s$ fields increase with increasing
spin. The most likely scenario is that there exists a $s_{\max}$ above which
all fields are inert. Without further calculations this $s_{\max}$ may be any
half-integer or integer between $1$ and $\infty,$ only future calculations can
resolve this problem.

The new SLFT setting of models involving $s\geq1$ fields leads also to new
problems which affect our understanding of symmetries. The standard view is
that inner symmetries encode the existence of superselection sectors of
observables \cite{Haag}. This works in both ways; if one starts from a model
with an inner symmetry one may construct its observable algebra as the fixed
point algebra under the action of the symmetry; vice versa one may recover the
field algebra from the smaller observable algebra by looking for all local
equivalence classes of its localizable representations.

With any such QFT with inner symmetries one may associate another one with the
same type of fields but changed couplings between them and possibly different
masses. In this way a symmetric QFT can be associated with models with the
same field content but broken symmetries. This conceptual situation changes in
the presence of $s\geq1.$ In the positivity maintaining SLFT setting the
presence of a second order induced $A\cdot A\left\vert \varphi\right\vert
^{2}$ term in scalar QED is the result of the causal localization principle of
QFT and does not require any gauge theoretical $\partial\rightarrow
D=\partial-iA~\partial$. Similarly the Lie structure of the $f_{abc~}%
$in$~L=f_{abc}F^{a,\mu\nu}A_{\mu}^{b}A_{\nu}^{c}+...$is a consequence of the
second order pair condition which in turn results from the implementation of
causal localization in a positivity-maintaining Hilbert space setting. 

Whereas inner symmetries can be (spontaneous or explicitly) broken, a QFT of
self-interaction vector mesons without the Lie structure of its couplings
would contradict the causal localization principles. One can also paraphrase
this situation by saying that there exists a quantum fibre bundle-like
structure which is totally intrinsic (i.e. not the consequence of quantizing a
classical structure). 

This remains somewhat hidden in GT where the BRST gauge symmetry formalism
results from a quantum adjustment of the fibre bundle properties of classical
gauge theory (Stora). Gauge theory accounts for gauge invariant local
observables and the S-matrix, but it misses to explain the properties of
interacting $s\geq1~$QFT as consequences of the causal localization
principles. A "gauge principle" bears no relation to the foundational concepts
of QFT, one only needs it in order to extract a physical subtheory from a
description which violates quantum theory's positivity property (which insures
its probability interpretation). 

In the next section we will address the important problem of higher spin
energy-momentum tensors. Even if fields turn out to be inert, they still
interact with classical gravity and modify the gravitational field through
Einstein-Hilbert back-reactions.

\section{The problem of the $s\geq2$ E-M tensor, coupling to gravity}

The classical energy-momentum tensor $T_{\mu\nu}^{cl}$ is a trace- and
divergence-less quadratic expression in terms of classical fields which can
conveniently be obtained within the Lagrangian formalism. For low spin
$s\leq1$ fields the Lagrangian quantization leads to the same free fields and
energy momentum tensor as that obtained in Wigner's representation theoretic
quantum setting.

However for $s\geq2$ the quantum free fields start to differ from their
classical counterparts; pl tensor potentials and their fermionic counterparts
are not solutions of Euler-Lagrange equation and the quantization leads to a
formalism which requires the use of indefinite metric Krein spaces. The
classical gauge symmetry ("local symmetry" in difference to global inner
symmetries) looses its physical content and becomes a formal device which
filters a physical subtheory (local observables, S-matrix). The form of the
energy-momentum (E-M) tensor in the positivity preserving description based on
Wigner%
\'{}%
s unitary representation theory turns out to be the same as that in the
indefinite metric setting; but this ceases to be the case for massless
$s\geq2$ fields.

For $m=0,~s=1$ the E-M tensor is the well-known expression%
\begin{equation}
T_{\mu\nu}^{P}~\simeq F_{\mu\kappa}F_{\nu}^{\kappa}-\frac{1}{4}\delta_{\mu\nu
}F_{\kappa\lambda}F^{\kappa\lambda}%
\end{equation}
It is easy to see that its massive counterpart is different by additional
$m^{2}$ contributions%

\begin{equation}
T_{\mu\nu}^{P}~\simeq F_{\mu\kappa}^{P}F_{\nu}^{P,\kappa}-\frac{1}{4}%
\delta_{\mu\nu}F_{\kappa\lambda}^{P}F^{P,\kappa\lambda}+g_{\mu\nu}%
m^{2}aA_{\kappa}^{P}A^{P,\kappa}+m^{2}bA_{\mu}^{P}A_{\nu}^{P} \label{s=1}%
\end{equation}
where the superscript $P$ refers to the pl Proca potentials. The conservation
of the E-M tensor uses the "massive Maxwell" equation%
\begin{align}
\partial^{\nu}T_{\mu\nu}^{P}  &  =F_{\mu}^{P\kappa}\partial^{\nu}F_{\nu\kappa
}^{P}\label{cyc}\\
\partial^{\nu}F_{\kappa\nu}^{P}  &  =m^{2}A_{\kappa}^{P}\rightarrow0\text{
}for~m\rightarrow0\nonumber
\end{align}
The remaining terms are compensated with the last 2 terms in (\ref{s=1}) if
one chooses $b=-2a$ and $a=1/2$;~all this is well-known.

Naively one would expect that in the massless limit these $m^{2}$ terms drop
out so that the massive tensor converges against the Maxwell E-M tensor. But
this does not happen, rather these terms contribute the E-M tensor of a pl
scalar massless field so that the degrees \footnote{This interesting remark I
owe to K.-H. Rehren.}. In fact besides the zero mass limit of sl vector
potential which describes a helicity $h=1$ situation which accounts for two
$h=\pm1$ degrees of freedom there is a massless pl scalar field $\phi$ defined
as $\lim m\phi(x,e)=\phi(x)~$which takes care of the remaining of the three
$s=1$ degrees of freedom.

This means that, different from the gauge formalism were the massless
indefinite metric vector potential accounts for only the two physical
(helicity) degrees of freedom, the positivity preserving SLFT also maintains
the degrees of freedom. The important role of the degrees of freedom
preserving escort field in the "fattening of the photon" (the "switching on
the mass" as the inverse of taking the massless limit) was already mentioned
in section 3.

There is no problem to extend this construction of conserved tensors quadratic
in the massive spin $s=n>1$ fields. For convenience we use the following
condensed index notation%
\begin{equation}
F_{\mu\kappa,\times}^{P}=as\ \partial_{\mu}A_{\kappa,\times}^{P},\text{
}\times=\kappa_{1},..\kappa_{n-1}\label{f}%
\end{equation}
where $A_{\kappa,\times}^{P}$ is a point-local totally symmetric $s=n~$tensor
potential and the anti-symmetrization refers to $\mu.\kappa$. Note that apart
from the case $s=1$ the $F$ is not the pl field strength; the latter
corresponds to the tensor $\mathcal{F}$ of degree $2s$ in (\ref{F}). The
dimension in mass units ("engineering" dimension) is $d_{en}=1$ for the
potentials $A^{P}$ and $2$ for the $F^{P},$ whereas their short distance
dimension increases as $d_{sd}=s+1$ for the $A^{P}$ as well as for the $F^{P}$
(the anti-symmetrization undoes the gain from the differentiation). The
differentiations which convert $A^{P}\ $into the field strength $\mathcal{F}$
(\ref{F}) increase the engineering dimension $d_{en}=d_{sd}=2s.$ The
$\mathcal{F}$ are the only $s\geq1$ pl fields which have a massless limit.

The construction of the E-M tensor $T_{\mu\nu}^{P}$ in terms of the $F^{p}$
(\ref{f})~for general massive$~s>1$ fields proceeds in the same way as
(\ref{s=1}). We again view $A_{\kappa,\times}^{P}$ as a $\times$-indexed
1-form, so that the exact 2-form $F_{\mu\kappa,\times}$ satisfies the cyclic
relation which implied the kinematic relation (\ref{cyc}). The divergence of
$F$ can be computed using the properties of the symmetric trace- and a
divergence- less $A^{P}$ tensor.

A conserved $s>1$ tensor which extends (\ref{s=1}) is%

\begin{equation}
T_{\mu\nu}^{P}=F_{\mu\kappa,\times}^{P}F_{\nu}^{P,\kappa,\times}-\frac{1}%
{4}g_{\mu\nu}F_{\lambda\kappa,\times}^{P}F^{P,\lambda,\kappa}+g_{\mu\nu}%
m^{2}\frac{1}{2}A_{\kappa,\times}^{P}A^{P.\kappa,\times}-m^{2}A_{\mu,\times
}^{P}A_{\nu}^{P.\times}\label{before}%
\end{equation}
Note that the $s$-dependent short distance dimension $d_{sd}=2+2s$ of $T^{P}$
is larger than its mass- ("engineering"-) dimension which remains at $d=4.$
But since in the calculation of the global Poincar\'{e} "charges" the spatial
integration together with the summation over the $\times$ indices leads to
cancellations of $m^{2}$-factors in the denominator against $p^{2}$-factors in
the numerator this discrepancy can be shown to disappear in the infinite
volume limit. \ 

The construction of the string-local E-M tensors follows similar formal
arguments. Instead of $A^{P}$ one starts from the string-local potential
$A_{\mu,\times\text{ }}$with $d_{sd}=d_{en}=1$ which was obtained by
successively "peeling off" leading short distance dimensions (\ref{ten})$.$The
differential form arguments (i.e. the use of the homogenous Maxwell relation)
is analogous. The divergence of $F$ is again of the form of $m^{2}B_{\mu}$
where $d_{sd}(B)=2$ and $B$ can be expressed in terms of $A$ and the $\phi$
(\ref{ten}). Hence the compensating quadratic $B$ terms are of a similar
algebraic form as before i.e.%
\begin{equation}
T_{\mu\nu}(x,e)=F_{\mu\kappa,\times}F_{\nu}^{\kappa,\times}-\frac{1}{4}%
g_{\mu\nu}F_{\mu\kappa,\times}F^{\mu\kappa,\times}+g_{\mu\nu}m^{2}\frac{1}%
{2}B_{\mu,\kappa}B^{\mu,\kappa}-m^{2}B_{\mu,\times}B_{\nu}^{\times}%
\end{equation}
the only formal difference is the lower $d_{sd}=2$ which is a result of the
weakening from pl to sl localization.

The claim is now that the $pl$ and its $sl$ counterpart have the same
conserved global charges. Although the charges densities $T_{0\nu}^{P}$ and
$T_{0\nu}$ are different they only differ by boundary contributions which
vanish in the infinite volume limit. The proof would render this article
somewhat unbalanced and will be presented together with the construction of a
consered $T_{\mu\nu}$ for infinite spin in a separate paper. 

Here we will be satisfied with a presentation of the ideas in the context of
the simpler case of a conserved free electromagnetic current. The conserved sl
current of a charge-carrying $s=1$ field is related to its pl counterpart as
($A^{P},\phi$ complex, $A=A^{P}+\partial\phi,~\partial\cdot A^{P}=0$)%
\begin{align}
&  j_{\mu}(x,e)=iA_{\kappa}^{\ast}(x,e)\overleftrightarrow{\partial_{\mu}%
}A^{\kappa}(x,e)=j_{\mu}^{P}(x)+\partial_{\kappa}G_{\mu}^{\kappa}%
(x,e)+m^{2}i\phi^{\ast}\overleftrightarrow{\partial_{\mu}}\phi\\
&  with~~\partial^{\kappa}G_{\kappa,\mu}=iA_{\kappa}^{P\ast}%
\overleftrightarrow{\partial_{\mu}}\partial^{\kappa}\phi+h.c.=i\partial
^{\kappa}(A_{\kappa}^{P\ast}\overleftrightarrow{\partial_{\mu}}\phi
)+h.c\nonumber
\end{align}
and since a spatial divergence does not contribute to the charge%
\begin{equation}
\int j_{0}(x,e)d^{3}x=\int j_{0}^{P}(x)d^{3}x+\int\partial^{0}G_{0,0}%
(x,e)d^{3}x-m^{2}\int j_{0}^{\phi}d^{3}x
\end{equation}

The last step consists in realizing that on can use $\partial\cdot A^{P}=0$ to
convert the middle term into spatial divergence%
\begin{align}
\partial^{0}G_{0,0}(x,e)  & =i\partial^{i}A_{0}^{P\ast}\overleftrightarrow
{\partial_{i}}\phi+h.c.=boundar\not y\text{ }term\\
j_{0}^{P}(x)  & =j_{0}(x,e)+m^{2}j_{0}^{\phi}(x,e)+boundary~term
\end{align}
whereupon it can be disposed of in the infinite volume limit. The scalar
contibution has a massless limit since $\lim_{m\rightarrow0}m\phi
(x,e)=\varphi(x)$ exists as a massless pl field. The massless sl $j_{0}%
(x,e)~$current lives in the Wigner-Fock $h=\pm1$ helicity space whereas the
massless scalar $\varphi$ current accounts for one degree of freedom. Hence
the 2+1 $h=1~$helicity+scalar degrees of freedom precisely match the 3 spin
$s=1$ degrees of freedom.

This highlights the important role of the escort field $\phi$ in a positivity
maintaining setting. This property is lost in the gauge theory; the indefinite
metric and ghost degrees of freedom are no substitute for the degrees of
freedom maintaining escort fields.  

The expected result of the extension to $s>1~$should be obvious. In addition
to the massless pl scalar field from $\lim_{m\rightarrow0}m^{s}\phi(x,e)$
their are sl helicity $\left\vert h\right\vert =s-k$ potentals of degree $k$
from $\lim_{m\rightarrow0}m^{s-k}\phi_{\mu_{1}..\mu_{k}}(x,e)=\varphi_{\mu
_{1}..\mu_{k}}(x,e)~$fields which account for the preservation of degrees of
freedom in the massless limit.

The relation between $T_{\mu\nu}(x,e)~$and $T_{\mu\nu}(x)$ is analogous but
somewhat more involved\footnote{In fact it was the $s=1$ $T_{\mu\nu}$ for
which Rehren pointed out to me that the $m^{2}$ terms converge to a finite
nontrivial massless contribution.}. One of the complications is that there are
several conserved $T_{\mu\nu}.$ One may use the Noether theorem for
translations or for the full Poincar\'{e} symmetry. The one which one would
use for gravitational backreactions is the one which is defined in terms of
the $g_{\mu\nu}$ metric variation of the action. We conjecture that the
equivalence of their pl with their sl form up to boundary terms holds for each one.

These results have an interesting connection with the No Go theorem by
Weinberg and Witten \cite{W-W} \cite{Por}. These authors claim that there
exist no electric charge-measuring conserved currents for $s\geq1$ and no E-M
tensors for $s\geq1.$ It one adds pl than the theorem hold for the rather
trivial reason that their existence would violate positivity as a cinsequence
of the nonexistence of zero mass $s\geq1$ pl potentials. In those cases the
correponding sl tensors perfectly exist and are even more natural in the
massive case in the sense that their operator- and engineering- dimensions
agree. 

Massless \textit{infinite} spin representations are described in terms of
scalar string-local fields with transcendent two-point functions as in
(\ref{two}); the most general form of intertwiners can be found in \cite{MSY}.
In this case there are no tensor potentials from which one can form $T_{\mu
\nu}$;$~$one must be content with the unitary Wigner description in terms of
$2s+1$-component wave functions and operators.

Formally the infinite spin representation may be viewed as the $s\rightarrow
\infty$ limit at fixed Pauli-Lubanski invariant $\kappa^{2}=m^{2}%
s(s+1)$\textit{ }$\ $and since our string-local fields guaranty the existence
of smooth massless limits of correlation functions, the following strategy
suggests itself. The first step consists in carrying out the $2s~$tensor
contractions which results in the following representation%
\begin{equation}
T_{\mu\nu}(x,e)=\sum_{\sigma,\kappa}\int\int e^{i(p-q)x}u_{\mu\nu}%
^{(+,-)}(p,q;\sigma,\kappa)a^{\ast}(p,\sigma)a(q,\kappa)\frac{d^{3}p}{2p_{0}%
}\frac{d^{3}q}{2q_{0}}+fluct.~terms+h.c. \label{in}%
\end{equation}
where the $T_{\mu\nu}$ intertwiners $u_{+,-},u_{+,+}$ and their conjugates are
contracted sums over products of spin $s$ tensor intertwiners.

The existence of a Pauli-Lubanski limit is connected with the limiting
behavior of these mass dependent composite $u(p,q)$ intertwiners for
$s\rightarrow\infty$ at varying mass $m^{2}=\kappa^{2}/s(s+1).$ One expects
that the infinite spin E-M tensor has the correct equal time commutation
relation with fields as (\ref{Ko}) or with the general class of such fields in
\cite{MSY}; since a computational check depends on explicit formulas for the
intertwiner in (\ref{in}) we defer the necessary calculation to a separate paper.

The construction of an E-M tensor is the sine qua non for a coupling of WS to
gravity; merely pointing to the energy positivity of the WS representation
does not address the important problem of backreaction \cite{DFP}.

The idea of a local generation of global spacetime symmetries receives support
from the algebraic setting of QFT where it can be shown that under very
general conditions ("split property") there exist a local implementation of
Poincar\'{e} transformations \cite{Haag}. In the point-local case this means
that the Poincar\'{e} transformation in a small parameter range around zero
acting on an operator localized in a compact double cone spacetime region can
be implemented by (highly non-unique) unitary operators which are localized in
a slightly larger region. For the noncompact localizable WS matter the region
would be a narrow spacelike cone whose core is a spacelike string.

\section{Concluding remarks}

The existence of forms of inert matter enriches the discussion about dark
matter in an interesting way. There are two problems with identifying both.
The inert WS matter seems to run into contradictions with astrophysical data
which show that dark matter hovers around galaxies in the form of a halo which
excludes "fleeting" massless matter\footnote{Assuming the noncompact massless
matter shares this property with photons.}. Inert massive matter can presently
not be excluded by astrophysical observations.

There is presently no serious problem with more conventional proposals which
identify dark matter with matter of low reactivity as WIMPS or Cold Dark
Matter. Such proposals become however increasingly problematic if refined
earthly detection attempts remain inconclusive and the failure of
counter-registering dark matter in terms of interactions with ordinary matter
forces darkness increaingly towards inertness ($\equiv$total darkness).

With refinements of astrophysical observations and the failure to see the
direct effects of dark matter in particle counters one is entering a "catch 22
situation": inert dark matter is consistent with its apparently exclusive
gravitational manifestations but gets into conflict with the role which
cosmologists attribute to it in the formation of ordinary matter in the Big
Bang standard model of cosmology.

Any particle counter observation of a new form of matter for which there are
good reasons to interpret it as a manifestation of the ubiquitous galactic
dark matter will eliminatete inert matter from the list of dark matter
candidates. But this would not diminish WS important role as a catalyzer of
new ideas concerning the interplay between Hilbert space positivity,
localization and short distance behavior for interactions involving higher
spin $s\geq1$ fields.

Independent of the problem of inert matter and its possible relation with dark
matter, Wigner's 1939 discovery of the WS representation class of the
Poincar\'{e} group and the more than 7 decades lasting attempts to understand
its causal localization properties are an important achievement n unravelling
unknown regions of QFT which remains our most successful theory about Nature's
physical properties.

\textbf{Acknowledgement}: I thank Jakob Yngvason and Jens Mund for their
continued interest in the issue of sl fields associated to the third Wigner
class whose construction was the result of a past joint collaboration
\cite{MSY}.

I am indebted to Christian K\"{o}hler for sending me a copy of his thesis
before publication. Last not least I thank Jos\'{e} Gracia-Bondia for a
critical reading of the manuscript. Last not least I am thankful to Henning
Rehren for sharing with me his interesting observation how the for spin
$s\geq1~$more natural higher spin string-local energy-momentum tensor
preserves the $2s+1$ degrees of freedom in the massless helicity limit.

\end{document}